\documentclass[twocolumn,showpacs,amsmath,amssymb,nofootinbib,prc]{revtex4-1}

\usepackage{graphicx} 
\usepackage{dcolumn}

\begin{document}
\title{Evolution of the $N=50$ gap from $Z=30$ to
$Z=38$ and extrapolation towards $^{78}$Ni.
}
\author{M.-G. Porquet$^1$}
\author{O. Sorlin$^2$}
\affiliation{
$^1$CSNSM, CNRS/IN2P3 and Universit\'e Paris-Sud, B\^at 104-108,
F-91405 Orsay, France\\
$^2$GANIL, CEA/DSM~-~CNRS/IN2P3, BP 55027, F-14076 Caen Cedex 5, France
}
\date{\today}
\begin{abstract}
The evolution of the $N=50$ gap is analyzed as a function of the
occupation of the proton $\pi f_{5/2}$ and $\pi p_{3/2}$ orbits. It is
based on experimental atomic masses,
using three different methods of one or two-neutron separation
energies of ground or isomeric states. We show that the effect of
correlations, which is maximized at $Z=32$ could be misleading with respect to the 
determination of the size of the shell gap, especially when using
the method with two-neutron separation energies. From the methods
that are the least perturbed by correlations, we estimate the
$N=50$ spherical shell gap in $^{78}_{28}$Ni$^{}_{50}$. Whether
$^{78}$Ni would be a rigid spherical or deformed nucleus is discussed
in comparison with other nuclei in which similar nucleon-nucleon
forces are at play.
\end{abstract}


\maketitle

\section{Introduction}

The persistence of the magic numbers remained a dogma for several
decades until the growing possibilities of
exploring nuclei far from stability have revealed that several magic
shell gaps were  fragile. In particular, the neutron-rich nuclei
$^{12}$Be$_{8}$, $^{32}$Mg$_{20}$, and $^{42}$Si$_{28}$ were found to
exhibit large collectivity in spite of their "magic" neutron numbers
N=8, 20, and 28 (see for instance~\cite{iw00,ga84,ba07}). This
disappearance of  traditional
magic numbers was ascribed to so far unexplored nuclear forces which
act to reduce the size of the spherical shell gaps~\cite{ot05}. Consequently 
nucleon excitations across these reduced gaps become
easier, increasing the amount of multi-particle multi-hole configurations.
Nuclei exhibiting these large correlations are usually deformed.

We have recently reviewed the major structural features along the
isotonic and isotopic chains around the spherical magic
numbers~\cite{so08}. By the way we have pointed out the role of
spin-flip interactions between protons and neutrons to modify the
harmonic oscillator or spin-orbit magic shells in a consistent
manner (see the various examples displayed in Figs.~45 and 46
of Ref.~\cite{so08}). We surmised in particular that nuclear forces
were acting to reduce the $N=50$ gap when $Z$ decreases towards $Z=28$. If true this 
would jeopardize the magicity of $^{78}_{28}$Ni$_{50}$. On the
other hand, recent values of atomic masses measured with high
precision using Penning trap mass spectrometers~\cite{ha08,ba08} were
interpreted in favor of a reinforced rigidity of the $N=50$ gap towards
$^{78}$Ni.

The aim of the present work is to present a detailed discussion on
the evolution of the $N=50$ gap from $Z=28$ to $Z=38$,
which includes the new high-precision mass measurements. We show in
particular the sensitivity of the obtained conclusions with the
methods used to derive the shell gap, taking the one or two neutron
separation energies from ground or isomeric states. We propose an
interpretation of
these discrepancies, and extrapolate the size of the $N=50$ gap to
$^{78}$Ni using what we consider to be the least biased method.

\section{The $N=50$ gap extracted from the properties of the $N=51-50-49$
isotones}\label{1neutron}

Following Ref.~\cite{so08}, we use the binding energy of the last
nucleon as a single-particle energy\footnote{Such a relation means that
the binding energy of the last nucleon, noted $BE_{1n}$(N) for the neutron,
is a negative quantity, i.e. the inverse of the separation energy,
noted $S_{1n}$(N).}
to study the evolution of the
two orbits $g_{9/2}$ and $d_{5/2}$ between which the $N=50$
spherical gap is formed. The Appendix of Ref.~\cite{so08} has warned
the reader about the limitations of such a method, as for
isotopes and isotones having a doubly-magic $N=Z$ core or for
non-rigid nuclei. In the latter case, the total binding energies of the
ground states do contain some correlation energies (from vibrational
motion, for instance) which shift the value of the binding energy of the
last nucleon.

\begin{figure}[!h]
\includegraphics[width=5.0cm]{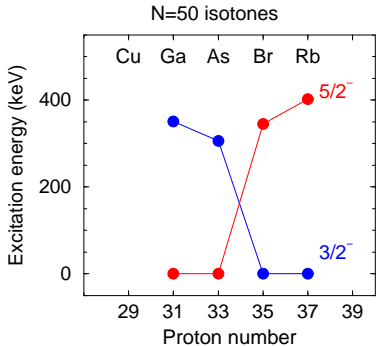}
\caption{(Color on line) Energy of the two first states of the $N=50$
isotones having an
odd-$Z$ value~\cite{NNDC}.}\label{oddZ}
\end{figure}

Between $Z=28$ and $Z=38$, protons occupy the
$\pi f_{5/2}$ and $\pi p_{3/2}$ orbits. Therefore the evolution of
the size of the $N=50$ shell gap depends on proton-neutron
interactions between the protons in the $\pi f_{5/2}$ and $\pi
p_{3/2}$ orbits and the neutrons in the $\nu g_{9/2}$ and $\nu
d_{5/2}$ orbits. The low-energy structure of the $N=50$ isotones
having odd-$Z$
values indicates that the $\pi f_{5/2}$  and $\pi p_{3/2}$ orbits
are very close to each others as the 3/2$^-_1$
and 5/2$^-_1$ states lie close in energy (see Fig.~\ref{oddZ}).
Therefore it is reasonable
to assume that the two proton orbits are filled simultaneously
within the $Z=28-38$ range and that the proton-neutron monopole
interactions at work are \emph{averaged} values between
$V^{pn}_{f5-g9}$ and $V^{pn}_{p3-g9}$ below $N=50$ (denoted as
$\bar{V}^{pn}_{f5/p3-g9}$), and between $V^{pn}_{f5-d5}$ and
$V^{pn}_{p3-d5}$ above $N=50$ ($\bar{V}^{pn}_{f5/p3-d5}$).

\begin{figure}[!h]
\includegraphics[width=5.5cm]{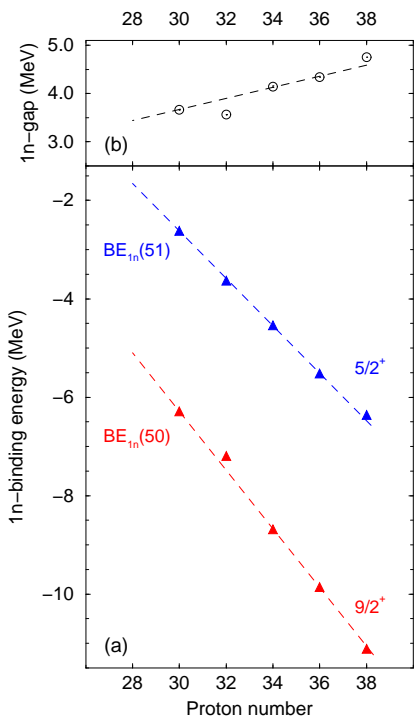}
\caption{(Color on line) (a) Experimental binding energies of the states 9/2$^+$
(5/2$^+$) located below (above) the $N=50$ magic number. The
experimental values of the atomic masses are from
Refs.~\cite{ba08,ha08,ame03}. The uncertainties are smaller than the
symbols. (b) Difference in binding energies of these two states
surrounding the gap at $N=50$. The dashed lines are the linear fits using the data at
$Z=30,34,36$ (see text).} \label{N50_S1n}
\end{figure}

The new high-precision atomic masses of
$^{83,82,81}_{~~~~~~32}$Ge~\cite{ha08} and
$^{81,80,79}_{~~~~~~30}$Zn~\cite{ba08} allow to display the binding
energies of the last neutron in the $N=51$ and $50$ isotones in
Fig.~\ref{N50_S1n}(a). This figure improves and enlarges
the Fig.~28 of Ref.~\cite{so08}.
We observe a quasi-linear evolution of the $BE_{1n}$ values, with
the exception of the $BE_{1n} (50)$ value at $Z=32$.
We discuss below the origin of this large singularity at $Z=32$.

The binding energy of the last neutron in $^{82}_{32}$Ge$_{50}$ lies
about 300~keV above the linear fit [see the red triangles and the red
dashed line in Fig.~\ref{N50_S1n}(a)]. This leads to a singularity at $Z=32$ which
might be viewed as arising from the presence of a subshell closure.
Under this assumption two slopes in the binding energies would be observed, one corresponding
to the filling of the $\pi p_{3/2}$ orbit until
$Z=32$ and the other to the filling of the $\pi f_{5/2}$ orbit afterwards. However, both the observed
ordering of the two orbits for $Z \le 33$ and their near degeneracy (see
Fig.~\ref{oddZ}) are strong arguments against the presence of such a
sub-shell closure. Therefore it is reasonable to think that this
singularity at $Z=32$ originates from the presence of deformation.
The $BE_{1n}(50)$ value for $^{82}$Ge involves the binding energies of
the $^{81,82}_{~~~32}$Ge$_{49,50}$ nuclei. In order to determine whether
$BE_{1n}(50)$ contains some amount of correlation energy, the structure
of $^{81}_{32}$Ge$_{49}$ has to be examined. The ground states of the
$N=49$ isotones are $9/2^+$, likely originating from a $(\nu
g_{9/2})^{-1}$ spherical configuration. The first excited states of
the $N=49$ isotones with Z$\ge$ 34 has I$^\pi$=1/2$^-$, it is likely to
be of $(\nu p_{1/2})^{-1}$ origin. The excitation energies of the $1/2^-$ states 
are slightly reduced with decreasing Z, i.e. 389~keV in $^{87}_{38}$Sr, 
305~keV in $^{85}_{36}$Kr and 228~keV in $^{83}_{34}$Se. When extrapolated to
$^{81}_{32}$Ge, the excited 1/2$^-$ state would be expected close to the 
9/2$^+$ ground state. Surprisingly its energy is as large as 
895~keV~\cite{ho81}. This suggests that the 9/2$^+$ ground state contains
some additional correlation energy which
decreases the atomic mass of $^{81}_{32}$Ge, shifting in turn the
binding energy of the last neutron in $^{82}_{32}$Ge. Discoveries of
the excited states built on the 'collective' 9/2$^+$ ground state and/or
measurements of the $E2$ transition probabilities
would ascertain the shape change of $^{81}_{32}$Ge as compared to
the heavier isotones. Unfortunately no excited state with I$^\pi >
9/2^+$ is known at the present time~\cite{NNDC}.

To perform the two linear fits drawn in Fig.~\ref{N50_S1n}(a), we have
chosen to eliminate the data at $Z=32$ and $Z=38$ as well, since $^{88}_{38}$Sr has the
main properties of a doubly-magic nucleus (it is used as an inert
core in many shell model calculations of nuclei with $Z,N \le 50$).
Indeed when a closed
shell is reached, a sudden reduction of correlation energies in principle occurs,
leading to a variation of the experimental binding energy of
the last neutron as mentioned in the Appendix of Ref.~\cite{so08}.
Thus taking into account the data at $Z=30$, 34, and 36, linear fits with
slopes of $|\bar{V}^{pn}_{f5/p3-g9}|=$~0.60~MeV and
$|\bar{V}^{pn}_{f5/p3-d5}|=$~0.48~MeV are obtained for $BE_{1n}(50)$
and $BE_{1n}(51)$, respectively.

The difference of the two binding energies, $BE_{1n}(51) - BE_{1n}(50)$,
gives the size of the correlated gap, which is drawn in
Fig.~\ref{N50_S1n}(b). The linear fit gives an extrapolated value at
$Z=28$ of 3.44~MeV. However this extrapolated
value will be enhanced if a reduction of correlation occurs at $Z=28$ as
it does at the $Z=38$ shell closure (the increase of the gap is 0.17~MeV here).
We shall come back to this point in Sect.~\ref{conclu}.
Before this, we compare the present conclusions obtained with
one-neutron binding energies to those derived when using two-neutron
binding energies, a method which is more extensively used in the literature.

\section{Evolution of the $N=50$ gap from the properties of the $N=52-50-48$
isotones}\label{2neutrons}
The behavior of the two-nucleon
separation energies is a widely used indicator of structural
evolution as for the emergence of magic numbers. For instance, the
two-neutron separation energies of each isotopic chain
display a sudden drop after the magic number when plotted as a
function of the neutron number, as for instance shown in the Fig.~2 of
Ref.~\cite{ha08}. Similarly, when plotted as a function of the proton
number, two neutron separation energies corresponding to each 
isotonic series form almost parallel and equidistant sequences,
displaying a sudden gap at the magic number. Such a plot is drawn in
the Fig.~3 of Ref.~\cite{ha08} where the evolution of the $N=50$
shell gap energy towards $^{78}$Ni is discussed.

\begin{figure}[!h]
\includegraphics[width=5.5cm]{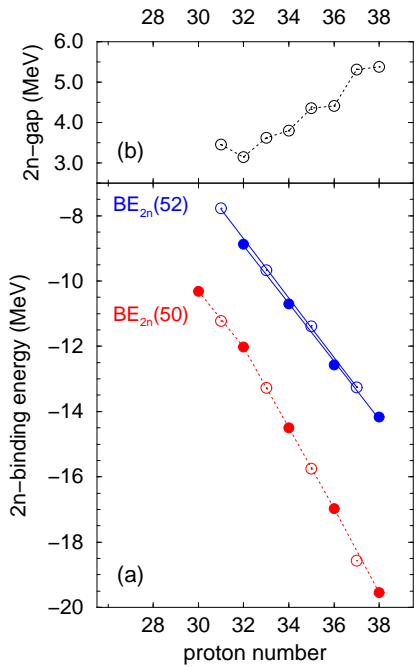}
\caption{(Color on line) (a) Experimental binding energies of the two last
neutrons in the $N=52$ isotones (blue circles) and the $N=50$
isotones (red circles). Experimental atomic masses are
taken from Refs.~\cite{ba08,ha08,ame03}. The uncertainties are
smaller than the symbols. (b) Difference of the two-neutron binding
energies between the $N=52$ and $N=50$ isotones.} \label{N50_S2n}
\end{figure}

We show in Fig.~\ref{N50_S2n}(a) the two-neutron binding energies using
the same conventions as those of Fig.~\ref{N50_S1n}
for the purpose of comparison.
When determining the evolution of
shell gaps using two neutron binding energies at $N=50$, the masses of
three isotopes, $N=48,50,52$, are involved. Assuming that there is no 
re-arrangement or collective excitation when adding or removing two 
neutrons from the semi-magic ($N=50$) cores, i.e. the $N=52$ and $N=48$ 
isotones are rigid spherical\footnote{Then their spectra would only display 
the states from the $j_n^2$ configuration, their spacings in energy being 
due to the two-body residual interactions.}, the $BE_{2n}$(52) and 
$BE_{2n}$(50) values are directly related to the binding energies of 
the $\nu d_{5/2}$ and $\nu g_{9/2}$ orbits, respectively.
Thus their variations are expected to be quasi-linear, the slope being
obtained from the average monopole proton-neutron interaction
involved, as discussed in Sect.~\ref{1neutron}. Moreover the
$BE_{2n}$ values should only depend on the occupation rate of the
$\pi f_{5/2}$  and $\pi p_{3/2}$ orbits, whatever the parity of $Z$
(even or odd). Figure~\ref{N50_S2n}(a) shows that the $BE_{2n}(50)$ trend
exhibits a kink at Z=32 and that the $BE_{2n}$(52) values obtained for 
odd-$Z$ and even-$Z$ nuclei do not belong to the same straight line, 
meaning that the criterion of rigid spherical shapes is not met.
As a result, the gap derived from
the difference of two-neutron binding energies, $BE_{2n}(52) -
BE_{2n}(50)$, shown in Fig.~\ref{N50_S2n}(b) displays a
complex behavior with (i) a staggering as a function of the parity of
$Z$ and (ii) a change of slope at $Z=32$. By pursuing $BE_{2n}(50)$ trend
from $Z=32$ to $Z=28$ one would find an increase of the $N=50$ gap
when reaching $^{78}$Ni~\cite{ha08}, contrary to what was derived in
Sect.~\ref{1neutron}.

\begin{figure}[!h]
\includegraphics[width=8.2cm]{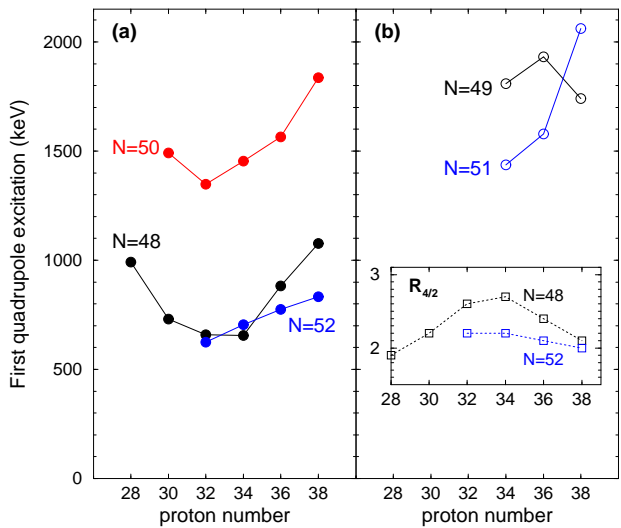}
\caption{(Color on line) Energy of the first quadrupole excitation built on the ground
state of the nuclei of interest: (a) The $2^+\rightarrow0^+_{gs}$
transitions of the even-even nuclei. (b) The
$13/2^+\rightarrow9/2^+_{gs}$ transitions of the $N=49$ isotones and
the $9/2^+\rightarrow5/2^+_{gs}$ transitions of the $N=51$ isotones.
The E(4$^+_1$)/E(2$^+_1$) ratios,
R$_{4/2}$, for the $N=52$ and $N=48$ isotones are shown in the
inset. The experimental information is from
Refs.~\cite{NNDC,va07,wi10,po09,po06}
} \label{firstexcitation}
\end{figure}

We suspect that extra nuclear correlations, which would be
maximized at $Z=32$, are responsible for this apparent change of
slope at $Z=32$. A good indicator of extra correlations can be revealed 
by the changes in the energy of first quadrupole
excitation built on the ground states. The experimental
results measured in the five isotonic chains ($N=48-52$) are used
for that purpose, namely (i) the energies of 
the $2^+_1\rightarrow0^+_{gs}$ transitions of the $N=48,50,52$ nuclei, (ii) the
energies of the $13/2^+_1\rightarrow9/2^+_{gs}$ transitions of the $N=49$ 
isotones, and (iii) the energies of the $9/2^+_1\rightarrow5/2^+_{gs}$ transitions 
of the $N=51$ isotones (see Fig.~\ref{firstexcitation}).

The 2$^+_1$ excitation of the semi-magic $N=50$ nuclei, which is due to their
proton configuration, lies between 1.35~MeV and 1.84~MeV 
[Fig.~\ref{firstexcitation}(a)]. The $N=51(49)$ isotones 
[Fig.~\ref{firstexcitation}(b)] display similar
values to those of their $N=50$ cores, that is expected if the neutron particle
(hole) is weakly coupled to the proton excitation of the core\footnote{Within
the weak-coupling scheme, the odd-$N$ nucleus exhibits a multiplet of states 
with spin values in the range [$|j_n-2|$, $j_n+2$] at an energy close to 
the 2$^+_1$ excitation of the core: This is well established in some of 
the $N=51(49)$ isotones (see Refs.~\cite{po06,po09} and references therein).}. 

On the other hand, the energies of 
the $N=52$ and $N=48$ isotones [Fig.~\ref{firstexcitation}(a)] are 
two times smaller than those of their $N=50$ cores.
Moreover, the quadrupole energy in the $N=48$ isotones has  a minimum 
at $Z=32-34$, indicating that the quadrupole deformation is maximum at 
mid proton
shell. This hypothesis is corroborated by the fact that the
E(4$^+_1$)/E(2$^+_1$) ratio reaches its maximum value there, $\sim$ 2.7,
which is intermediate between vibrator and rotor nuclei
[see the inset of Fig.~\ref{firstexcitation}(b)]. Thus extra binding 
energy due to
correlations depresses to a larger extent the atomic mass of the 
ground state of
$N=48$ isotones at the mid proton-shell,
giving rise to the observed increase of $BE_{2n}$(50). The same
mechanism has been recently discussed for isotopes with $N
\sim 90$ (see the Section~3 of Ref.~\cite{ca09}).

As for the $N=52$ isotones, their quadrupole energy decreases slowly 
[see Fig.~\ref{firstexcitation}(a)].
Even though their low 2$^+_1$ energies are not far from those of the $N=48$
isotones, their quadrupole collectivity is different as demonstrated by
the fact that the E(4$^+_1$)/E(2$^+_1$) ratios remain almost
constant and close to 2, that is the vibrational limit.  Since the
collective behavior of the $N=52$ isotones does not vary significantly with $Z$,
the extra binding energy depresses equally all the ground state masses,
thus explaining why the evolution of $BE_{2n}$(52) remains quasi linear.
One may assume that the extra binding is not the same for the even-Z
and odd-Z isotones, leading to the two straight lines observed in
Fig.~\ref{N50_S2n}.

In summary, the quadrupole characters associated to the low-energy
states of the $N=48$ and $N=52$ isotones add correlation energies to the 
$BE_{2n}$ values, which could therefore not be used to determine the
evolution of the spherical orbits bounding the $N=50$
gap. Such a statement had been already put on a firm
theoretical basis in several regions of the nuclear chart for which
static deformation and dynamic fluctuations around the mean-field
ground states have proven to modify the atomic masses and apparent
shell gaps (see for instance, Ref.~\cite{be08} and references therein).
To reduce the amount
of correlations in the determination of shell gaps, we propose to
use the binding energy of the 8$^+$ isomer of the $N=48$ isotones
instead of the one of their ground state.

\section{Evolution of the $N=50$ gap using the 8$^+$ isomeric state}\label{isomer}
Isomeric 8$^+$ states have been found around 3~MeV in all the
$N=48$ isotones. Their configuration was assigned to $(\nu
g_{9/2})^{-2}$ in which the two holes couple to the maximum spin
value J=8 in a spherical configuration (see
Ref.~\cite{po09} and references therein).
The isomerism is due to the relatively small energy between
the 8$^+_1$ and 6$^+_1$ states. Such a small energy difference implies that the 
8$^+$ isomers cannot belong to a vibrational/rotational band, and are well
separated in energy from any 8$^+$ 'collective' state. It follows that the 
configuration of these 8$^+$ isomers is assumed
to be rather pure, contrary to that of the ground state. The binding
energy of such a spherical state, being directly related to the one
of the orbit, gives another mean to characterize the evolution
of the $\nu g_{9/2}$ energy as a function of the proton number.

\begin{figure}[!h]
\includegraphics[width=8.2cm]{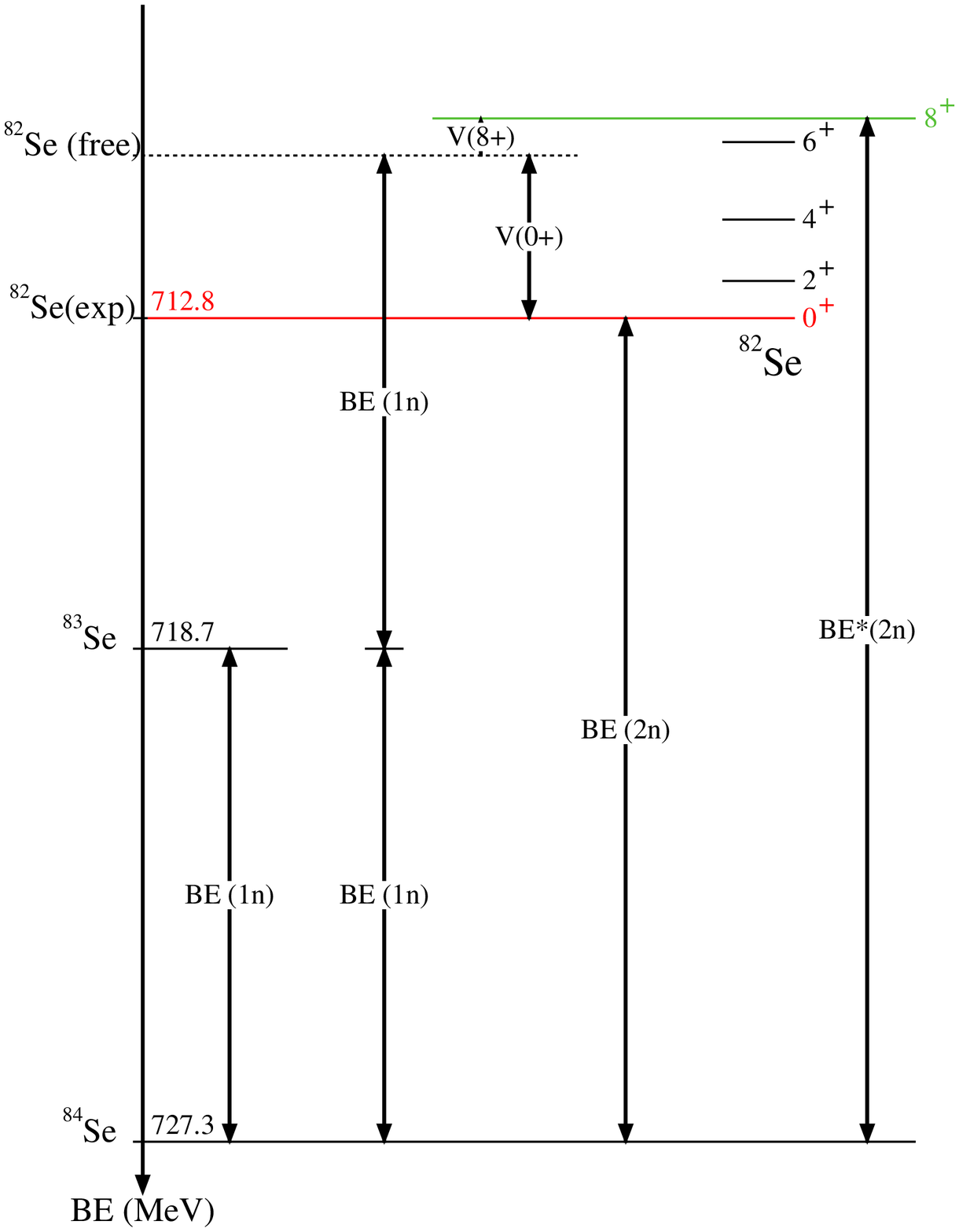}
\caption{(Color on line) Pictorial illustration of the total binding energies,
BE, and
various neutron binding energies, $BE_{1n}$, $2BE_{1n}$, $BE_{2n}$
et $BE^*_{2n}$ for the Se isotopes, showing the correlation energy of
the 0$^+_{gs}$ state, V(0$^+$), and the residual interaction of the
8$^+$ spherical state, V(8$^+$), in $^{82}$Se$_{48}$.}
\label{IntRes82Se}
\end{figure}

For this purpose,
we define another binding energy of the last two neutrons of the $N=50$
isotones, $BE^*_{2n}(50)$, using the total binding
energies (BE) of the nuclei of interest and the excitation energy of
the 8$^+$ isomeric states:
\begin{eqnarray}\label{S*2n}
BE^*_{2n}(Z,50) = -[BE_{gs}(Z,50) - BE_{8^+}(Z,48)]\nonumber \\
\equiv [BE_{gs}(Z,48) - E_{exc}(8^+)] - BE_{gs}(Z,50)
\end{eqnarray}
The meaning of the neutron binding energies, $BE_{1n}$,
$2BE_{1n}$, $BE_{2n}$ et $BE^*_{2n}$, is given in Fig.~\ref{IntRes82Se}
for the $^{82,83,84}$Se isotopes.

\begin{figure}[!h]
\includegraphics[width=5.5cm]{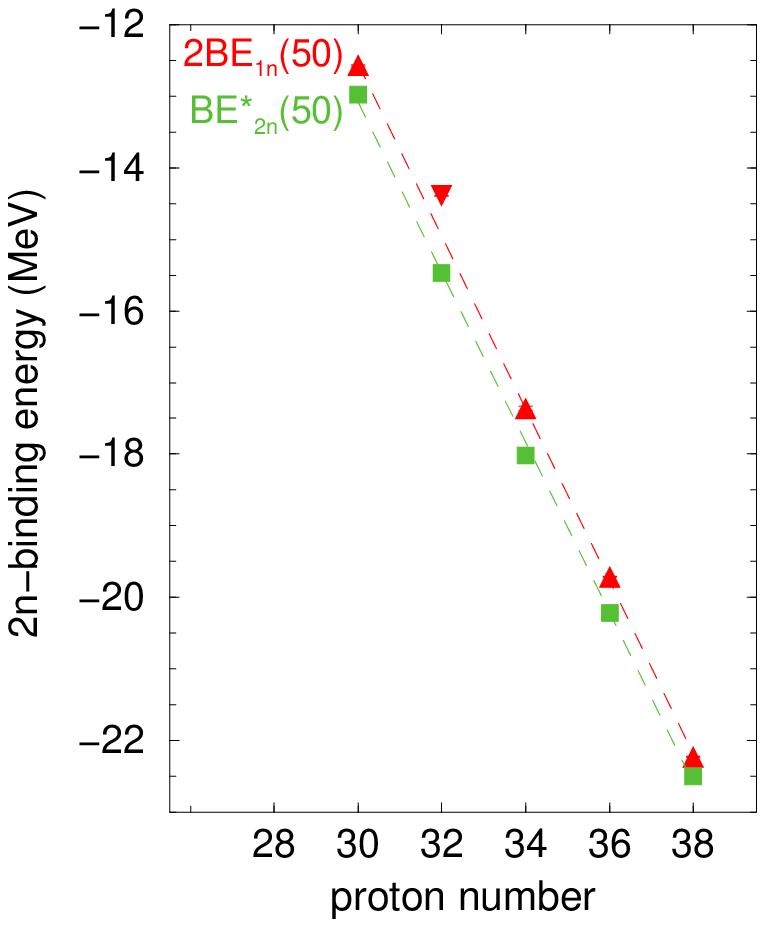}
\caption{(Color on line) Experimental binding energies of the two last neutrons in
$N=50$ isotones: Values of $2BE_{1n}$(50) (red triangles, cf.
Fig.~\ref{N50_S1n}) and of $BE_{2n}^*$(50) (green squares, cf.
Eq.~\ref{S*2n}). The difference between the two lines is, by
definition, the interaction energy between the two neutrons located
in the $\nu g_{9/2}$ orbit, which couple to J= 8$^+$. The
uncertainties are smaller than the symbols.} \label{S2n8plus}
\end{figure}

\begin{figure}[!h]
\includegraphics[width=5.5cm]{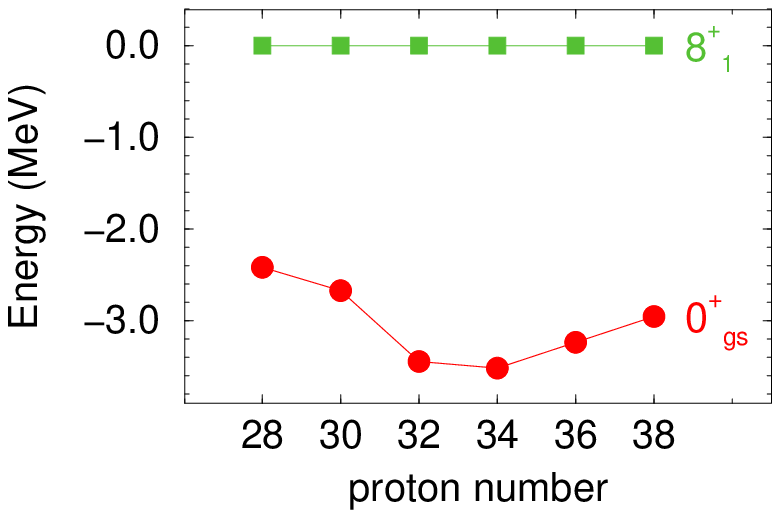}
\caption{(Color on line) Distance in energy between the 0$^+_{gs}$ and the 8$^+_{1}$
states for the $N=48$ isotones as a function of the proton number.
The energies of the 8$^+_1$ state are fixed to 0.} \label{8plusN48}
\end{figure}

The evolution of the $BE^*_{2n}$ values as a function of $Z$ is drawn
in Fig.~\ref{S2n8plus}.
Noteworthy is the fact that the change of
slope in $BE_{2n}(50)$ observed at $Z=32$ in Fig.~\ref{N50_S2n} has
disappeared. The $BE^*_{2n}$ values rather display a straight line
with a slope of -1.19(2)~MeV/Z. This slope matches the one of
-1.20(1)~MeV/Z obtained using the $2BE_{1n}$ values. We therefore
deduce that a large part of the deviation of $BE_{2n}(50)$ to a
straight line originates from the correlation energy of the
ground state of the $N=48$ isotones, labeled as V(0$^+$)
in Fig.~\ref{IntRes82Se}. Indeed the relative distance in
energy between the 0$^+_{gs}$ and 8$^+_1$ states for the $N=48$ isotones
is maximum at $Z
\simeq 32,34$ as shown in Fig.~\ref{8plusN48}.

In summary, the approaches using $BE_{1n}$ and $BE^*_{2n}$ give
similar results with respect to the behavior of the $\nu g_{9/2}$
orbit as a function of the occupation of the $\pi f_{5/2}$  and $\pi
p_{3/2}$ orbits. As these two approaches do not strictly use the
same experimental data (only the atomic masses of the $N=50$ cores
are in common), we have more confidence in the linear fits proposed
in Sect.~\ref{1neutron}.

\section{Discussion}\label{conclu}

The structure of the neutron-rich nucleus $^{78}$Ni, having $Z=28$
and $N=50$ magic numbers, depends on the size of the $N=50$ and
$Z=28$ spherical gaps and on the amount of the quadrupole correlations
provided by generating excitations across them\footnote{Quadrupole
correlations across the $N=50$ and $Z=28$ spherical gaps may be strongly
favored since the two orbits bounding both of them have $\Delta l =2$:
$\pi f_{7/2}$  and $\pi p_{3/2}$ in the one hand, $\nu g_{9/2}$ and
$\nu d_{5/2}$ in the other hand.}. In this section, we discuss the
behaviors of the two gaps in order to argue about the doubly-magic
nature of $^{78}$Ni.

\begin{figure}[!h]
\includegraphics[width=5.5cm]{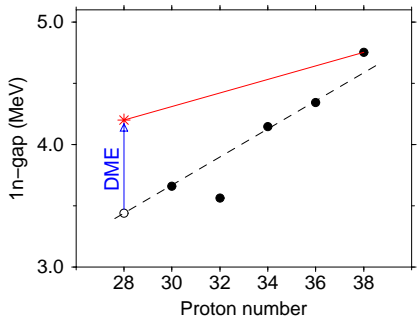}
\caption{(Color on line) Difference of the binding energies of the two states
surrounding the gap at $N=50$ [cf. Fig.~\ref{N50_S1n}(b)].
When using the averaged DME value of $^{48}$Ca (see text), the $N=50$ gap 
would amount to 4.2~MeV for $Z=28$. The solid line
gives the evolution of the $N=50$ gap from $Z=28$ to $Z=38$.}
\label{gapN50_discuss}
\end{figure}

In Sect.~\ref{1neutron}, we gave the extrapolation of the $N=50$ 
correlated gap at $Z=28$, 3.44~MeV, and  the reduction of correlation
energy due to the $Z=38$ shell closure, which amounts to 0.17~MeV.
Nevertheless, a much larger singularity may occur in a nucleon shell
gap when the number of nucleons of the other species is equal to a magic
number, the so-called "doubly--magic effect" (DME).
Besides the well-known cases dealing with the Wigner term which gives an
additional binding to nuclei having $N=Z$, there are a few other cases
which are shown in several figures of Ref.~\cite{so08}.
For instance, the $Z=20$ gap is enhanced by 0.86~MeV at $N=28$ (see the
Fig.~17 of Ref.~\cite{so08}) and the $N=28$ gap is enhanced by 0.71~MeV
at $Z=20$ (see the Fig.~20 of Ref.~\cite{so08}), that gives an averaged DME
value of 0.78~MeV. If the odd-$N$ nuclei close to $^{78}$Ni$_{50}$ behaves
as odd-A nuclei close to $^{48}$Ca, the value of the $N=50$ spherical
gap for $Z=28$ would be 4.2~MeV, as shown in Fig.~\ref{gapN50_discuss}.

The $Z=28$ spherical gap is reduced when going towards $N=50$.
This has been demonstrated by the monopole migration of the $\pi
f_{5/2}$ orbit as the $\nu g_{9/2}$ is filled~\cite{fr01},
provoking a inversion between the $\pi p_{3/2}$ and $\pi f_{5/2}$ in
the $_{29}$Cu chain at $A=75$~\cite{fl09}. More indirectly, the
$B(E2; 0^+ \rightarrow 2^+_1)$ excitation strengths measured in two
neutron-rich $_{28}$Ni isotopes~\cite{pe06,ao10} can be reproduced
only when invoking an increasing amount of proton core excitations
from the $\pi f_{7/2}$ orbital to the other $\pi fp$ orbits.
In addition, the rapid lowering of the first 1/2$^-$ state in $^{69-75}$Cu
and the enhancement of the $B(E2; 1/2^-_1 \rightarrow 3/2^-_1)$ transition
rates provide other proofs of the major role of the $\pi f_{7/2}$
orbital in the description of the $Z \geq 28$ isotopes~\cite{si10}.
Moreover, the SM calculations of Ref.~\cite{si10} indicate that the $Z=28$
shell closure gets reduced by about 0.7~MeV between $^{68}$Ni and $^{78}$Ni.

Theoretical models predict $^{78}_{28}$Ni$_{50}$ to be a doubly-magic
spherical nucleus (see, for instance, Refs.~\cite{be09,de10}).
Using the present empirical findings, one can only argue on
qualitative statements upon its magicity. The two magic numbers,
28 and 50, are created mainly by the spin-orbit interaction and
$^{78}_{28}$Ni$_{50}$ belongs to the same family of other doubly
spin-orbit nuclei as $^{20}_{~6}$C$^{}_{14}$, $^{42}_{14}$Si$^{}_{28}$,
and $^{132}_{~50}$Sn$^{}_{82}$. The latter has the major characteristics of a
doubly-magic spherical nucleus~\cite{fo94,jo10} while the two others 
have not~\cite{st08,ba07}. The first excited state of
$^{42}_{14}$Si$^{}_{28}$ measured at very low energy is due to the erosion
of both the $Z=14$ and $N=28$ shell closures (caused by action of the mutual
proton and neutron forces) \emph{and} to the quadrupole correlations
between states bounding the two gaps~\cite{ba07}. Such a deformed
configuration could be found for the ground state of $^{78}$Ni or a
low-lying state. Under this latter assumption a shape coexistence could be 
found for $^{78}$Ni, as it has been 
recently observed at the $N=28$ shell for $^{44}_{16}$S$^{}_{28}$~\cite{fo10}.

\section{Summary}\label{last}
The evolution of the $N=50$ gap between $Z=30$ and $Z=38$
has been studied by means of three different methods based on one or
two-neutron separation energies of ground or isomeric states, which all
take into account the newly determined atomic masses of Refs.~\cite{ha08,ba08}.
By extracting the $N=50$ gap from these methods, different degrees
of correlations are intrinsically involved. These correlations
distort the extracted gap value, which is rather then a correlated
one. The correlations are the strongest at $Z=32$, and weakest at
the two proton shell closures $Z=38$ and $Z=28$. It has been shown
that the use of two-neutron separation energies to analyze the evolution
of the $N=50$ gap value is the most subject to
correlations provided in particular by the $N=48$ nuclei. This is
partly remedied by using atomic masses of the least correlated
$8^+$ isomeric state of the $N=48$ nuclei, rather than those of
the $N=48$ ground states. Gathering the results of the three
methods, we propose a global reduction of the $N=50$ gap between
$Z=38$ and $Z=28$ by about 0.55~MeV, which is to combine to the
expected reduction of the $Z=28$ gap at $N=50$. It follows that
the structure of the $^{78}$Ni nucleus is probably intermediate
between the deformed $^{42}$Si and the spherical $^{132}$Sn.
Whether $^{78}$Ni would be deformed, spherical or exhibit
a shape coexistence depends on a delicate balance between the
size of the $Z=28$ and $N=50$ spherical gaps which preserve its
sphericity, and the amount of correlations brought by promoting
nucleons across these gaps leading to deformation.

\begin{acknowledgments}
We thank J.-C. Thomas for fruitful discussions.
\end{acknowledgments}

\end{document}